# Atomic cluster expansion interatomic potential for defects and thermodynamics of Cu-W system

Jiahao Pan[1], Huiqun Cheng[1], Gaosheng Yan[1], Lei Zhang[*,2], Wenshan Yu[*,1], Shengping Shen[*,1]

[1]State Key Laboratory for Strength and Vibration of Mechanical Structures, Shaanxi Engineering Laboratory for Vibration Control of Aerospace Structures, School of Aerospace Engineering, Xi'an Jiaotong University, Xi'an 710049, People's Republic of China.

[2]Engineering and Technology Institute Groningen, Faculty of Science and Engineering, University of Groningen, Nijenborgh 4, 9747 AG Groningen, The Netherlands.

[*]corresponding to: lei.zhang@rug.nl (L. Zhang) wenshan@xjtu.edu.cn (W.S. Yu) sshen@xjtu.edu.cn (S.P. Shen)

**Abstract**

The unique properties exhibited in immiscible metals, such as excellent strength, hardness, and radiation-damage tolerance, have stimulated the interest of many researchers. As a typical immiscible metal system, the Cu-W nano-multilayers combine the plasticity of copper and the strength of tungsten, making it a suitable candidate for applications in aerospace, nuclear fusion engineering, and electronic packaging *etc*. To understand the atomistic origin of the defects and thermodynamics of the Cu-W immiscible system, we have developed an accurate machine learning interatomic potential (ML-IAP) for Cu-W based on the atomic cluster expansion (ACE) method. The Cu-W ACE potential can faithfully reproduce the fundamental properties of Cu and W predicted by density functional theory (DFT). Moreover, the thermodynamical properties, such as the melting point, coefficient of thermal expansion, diffusion coefficient, and equation of the state curve of the Cu-W solid solution, are calculated and compared against DFT and experiments. Monte Carlo Molecular Dynamics (MC-MD) simulations performed with the Cu-W ACE potential predict the experimentally observed phase separation and uphill diffusion phenomena. Our findings not only provide an accurate ACE potential for describing the Cu-W immiscible system, but also shed light on understanding the atomistic mechanism during the Cu-W nano-multilayers formation process.

## I. INTRODUCTION

In most alloys, the heat of formation is negative, suggesting that the system naturally tends to form alloys at the atomic level. On the other hand, immiscible alloys have a positive heat of formation, indicating that they lack the thermodynamical driving force at ambient temperature and pressure to alloy directly[1]. With the development of physical mixing and deposition techniques, new immiscible alloys are generated, which exhibit excellent properties that simple metals do not possess. For example, Cu-X composite materials (X represents W, Mo, Ta, Nb, and other refractory metals) are widely used in aerospace, nuclear fusion engineering, electronic packaging, and anti-armor weapons because of their excellent hardness, strength, wear resistance, and corrosion resistance [2-6].

As a typical immiscible system, Cu-W nano-multilayers combine the strength,

hardness, corrosion resistance, radiation resistance of W and the thermal conductivity, electrical conductivity, plasticity of Cu [7]. The nano-multilayers containing 13.1 at. %W exhibit a hardness of 5.9 GPa, which is 55% higher than that calculated by the mixed rule [8]. The underlying mechanisms have not been investigated. A better understanding of the dynamical processes at the atomic scale is required, which is usually beyond the capability of experimental techniques.

Density functional theory (DFT) is based on quantum mechanics and is widely employed to compute properties and understand mechanisms at the atomic scale in the materials science community. DFT is capable of accurately predicting material properties with near quantum accuracy. However, DFT can only be applied to small times and lengths due to computational resource limitations (typically hundreds of atoms and less than 1 ns) [9]. Instead, classical molecular dynamics (MD) can be used to perform atomistic simulations with length and time scales that are orders of magnitude larger than DFT. The quality of MD simulations depends on the accuracy of the interatomic potentials (IAPs). The available IAPs for Cu-W systems are mostly empirical potentials. Gong [10] developed an embedded-atom method (EAM) potential based on the first-principle predicted lattice constant and binding energy. The proposed EAM [10] predicts that the intrinsic glass formation range of the Cu-W system is consistent with experimental observation. By considering the lattice constant and cohesion energy of $Cu_3W$ compounds, Wei [11] *et al*. further modified the EAM proposed by Gong *et al*. The modified EAM is able to reproduce the mechanical properties and enthalpy of formation of some solid solutions while the EAM potentials proposed by Gong [10] and Zhou [12] are unable to predict those properties. Recently, Wang [13] *et al.* developed an EAM potential to study the diffusion behavior of Cu-W interfaces with various crystallographic orientations. Yang [14] *et al.* optimized the EAM [10] by additionally fitting the structural energy difference between face-centered cubic (fcc) and body-centered cubic (bcc) structures of W and Cu. However, the modified EAM fails to predict the heat of formation for fcc solid solutions with low Cu concentration [14].

While the classical potentials are limited by the fixed functional form and few fitting parameters, Machine Learning (ML) IAPs have demonstrated significant advancements, achieving near-DFT accuracy in predicting a wide range of properties for various systems [15-18]. Among all the popular ML-IAPs, the atomic cluster expansion (ACE) method has proved to be one of the most efficient and accurate machine learning potentials [19,20]. Here, we develop an ACE potential for the Cu-W system to investigate the defects and thermodynamical properties of the Cu-W nano-multilayers. The remainder of this paper is structured as follows. In Sec. II, we present the basic concept of the ACE method, the details of database creation, and training the ACE potential. In Sec. III, we comprehensively assess the accuracy of the ACE potential for essential and thermodynamical properties, including elasticity, surface energy, phonon dispersion, melting temperature, coefficient of thermal expansion, diffusion coefficient, generalized stacking fault energy, work of separation, and Cu-W formation energy. Moreover, we reproduce the experimental phase separation of Cu-W nano-multilayers with MC-MD atomistic simulations based on the Cu-W ACE potential.

The main findings are summarized in Sec. IV.

## II. METHODS

### A. Atomic structures of the DFT database

ML potentials require an accurate and comprehensive database which is usually constructed via quantum mechanical calculations [21,22]. In general, the database should contain possible phases of the material at various temperatures and pressures. The defect structure relevant to the properties that the potential aims to predict should also be included [23,24]. Here, we first generated various bulk crystal configurations for Cu and W, as shown in Table I [25]. Initial configurations are taken from the Material Project, including various phases of Cu (Cubic (Fm$\bar{3}$m, Im$\bar{3}$m), Hexagonal (P6$_3$/mmc), Tetragonal (I4/mmm), Orthorhombic (Cmcm), Trigonal (P$\bar{3}$m1), Monoclinic P12/m1) and W Cubic (Im$\bar{3}$m, Pm$\bar{3}$n, Fm$\bar{3}$m), Orthorhombic (Pbcm), Hexagonal (P6$_3$/mmc) Tetragonal (I4$_1$/amd) Monoclinic (P12/m1)) [26]. We also incorporated low-index surfaces and some commonly observed grain boundary configurations (Σ3, Σ5, Σ7, Σ9). To characterize the Cu-W solid solution, the configurations with varying element ratios in both the fcc and bcc phases are included. This approach guarantees a comprehensive representation of the Cu-W solid solution across different phases. Moreover, a set of nonequilibrium configurations is generated by adding random noise (within ±5%) to the volumes and atomic positions of the equilibrium configurations. The energies and forces of each configuration are calculated using the Vienna ab initio simulation package (VASP) [27-29]. The PBE functional is used to approximate the exchange and correlation energy [30]. The cutoff for plane wave energy is 500eV. We use Vaspkit [31] software to generate Γ-centered k-points automatically with a k-mesh spacing of 0.015 Å$^{-1}$ for the primitive cells and 0.03 Å$^{-1}$ for the other configurations.

TABLE I. Configurations in the database.

| Database | | Compositions | Temperature | Number of configurations |
|---|---|---|---|---|
| **Primitive cell** | | fcc, bcc, hexagonal close-packed (hcp), double hexagonal close-packed (dhcp), et al [32] | 0K | 3348 |
| **Surface** | Cu | (100),(110),(111),(210),(211),(221), (310),(311),(320),(321),(322),(332) [32] | 0K | 357 |
| | W | (100),(110),(111),(210),(211),(221), (310),(311),(320),(321),(322),(331), (332) [32] | | |
| **Grain boundary** | Cu | Σ3(1-1-2), Σ3(1-10), Σ3(111), Σ5(0-1-3), Σ5(0-21), Σ7(3-2-1) [32] | 0K | 122 |
| | W | Σ3(110), Σ3(111), Σ3(112), Σ5(012), Σ5(013), Σ5(100), Σ7(132), Σ9(110), Σ9(221) [32] | | |
| **Supercell** | Cu | NPT (vacancy, no defects) | 300K,1000K,2000K | 148 |
| | W | NPT (vacancy, no defects) | 300K,4000K,5000K | |
| **Solid solution** | | bcc ($Cu_xW_{1-x}$, x from 0.0625 to 0.9375), fcc ($Cu_xW_{1-x}$, x from 0.03125 to 0.96875) | 0K | 9770 |

**B. Atomic cluster expansion methodology**

The ACE methodology combines the advantages of ML methods and physically based models of atomic interactions. As demonstrated by the Pareto front, the ACE potential strikes a balance between accuracy and efficiency, and has good extrapolation performance [19]. Here, we provide a concise summary of ACE[19,33-38].

ACE provides a complete and hierarchical basis functions to represent atomic property

$$\varphi_i^{(p)} = \sum_v^{n_v} c_v^{(p)} B_{iv}, \qquad (1)$$

where $B_{iv}$, $c_v^{(p)}$ are the basis functions constructed by ACE and the expansion coefficients, respectively. In the training, we use the implementation that the local energy is decomposed to two atomic properties with a Finnis-Sinclair square-root embedding

$$E_i = \varphi_i^{(1)} + \sqrt{\varphi_i^{(2)}}. \qquad (2)$$

The loss function $L$ is given by

$$L = (1-\kappa)\Delta_{Energy}^2 + \kappa\Delta_{Force}^2 + \Delta_{Coeff} + \Delta_{Rad}, \qquad (3)$$

where $\Delta_{Energy}^2$ and $\Delta_{Force}^2$ are the mean square error of the difference between energies and forces predicted in DFT and ACE potential. $\Delta_{Coeff}$ and $\Delta_{Rad}$ are used to smoothen the energy hypersurface and radial functions respectively. $\kappa$ balances energy and force contribution in the loss function.

### C. Fitting ACE potential

The ACE parameterization is carried out using the *pacemaker* package [36]. We found that the energy of a (Cu/W) dipole varies within 2 meV/atom once two atoms were separated by more than 7 Å. Therefore, the cutoff radius was set as 7 Å. It has been shown that hierarchical schemes have better performance than a single fit on energy and force root mean square error (RMSE) values [36]. We thus adopted a hierarchical power-order scheme to sequentially add ACE basis functions during the fitting process. The structures with energies higher than the ground-state energy of 15 eV/atom or forces higher than 30 eV/atom were removed from the database during training. Higher weights were given to structures within 4 eV/atom above the convex hull to better capture configurations far from equilibrium, thereby enhancing the transferability of the potential. In the loss function, κ was set to 0.01, indicating that the weights of energy and force were set to 0.99 and 0.01, respectively. Figure 1 plots the energy and force RMSE of the Cu-W ACE potential versus the reference DFT data for the training and testing datasets. The RMSE values for energy (forces) are 6.21 meV/atom (176.49 meV/Å) and 4.36 meV/atom (169.92 meV/Å) for the training and testing datasets, respectively. For the lower energy structures within the 4 eV above the convex hull, the RMSE values for energy (forces) are 2.56 meV/atom (77.37 meV/Å) and 2.71 meV/atom (69.44 meV/Å) for the training and testing datasets, respectively.

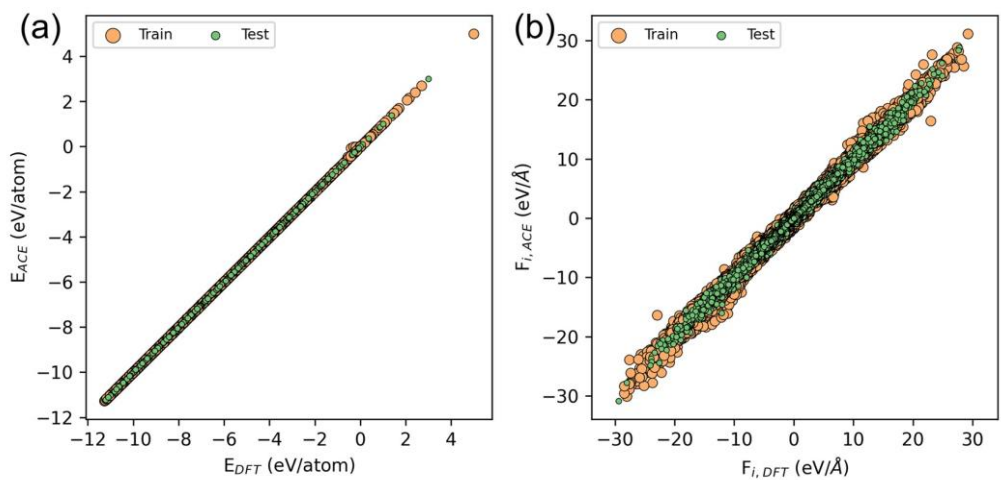

FIG. 1. Comparison of energies (a) and forces (b) predicted by ACE and DFT.

## III. RESULT & DISCUSSION

### A. Basic validation of the Cu-W ACE

A benchmark of various fundamental properties for elemental Cu and W is conducted, including the lattice parameter $a_0$, elastic constants ($C_{11}$, $C_{12}$, $C_{44}$), bulk modulus ($B$), vacancy formation energy ($E^f_{vac}$), and surface energies ($E^f_{(100)}$, $E^f_{(110)}$, $E^f_{(111)}$). The predictions are compared with DFT, as shown in Figure 2. A more detailed table is listed in Supplementary Material S1. The equilibrium lattice parameters predicted by the ACE potential are almost identical to DFT values. The relative error of elastic constants, bulk moduli, surface energy, and vacancy formation energy is within ± 6% compared to DFT except for the $C_{44}$ of W. More weights given to the nonequilibrium configurations during training may lead to deviations in predicting elastic constants. The orderings of surface energy from DFT predictions, i.e., $E^f_{(100)} > E^f_{(110)} > E^f_{(111)}$ for Cu and $E^f_{(100)} > E^f_{(110)} > E^f_{(111)}$ for W, are correctly reproduced by the ACE potential.

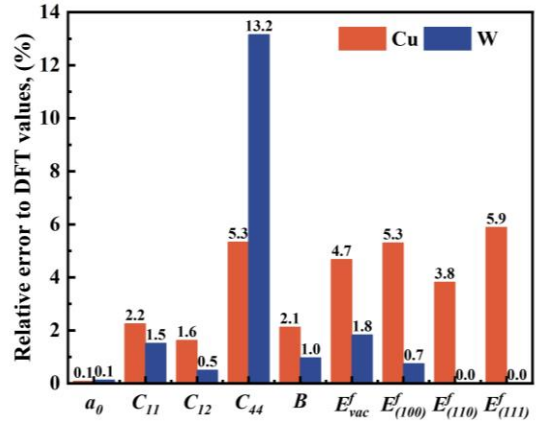

FIG. 2. Relative error of ACE potential prediction with respect to DFT calculations, including lattice constant ($a_0$), elastic constants ($C_{11}$, $C_{12}$, $C_{44}$), bulk moduli ($B$), surface energies ($E^f_{(100)}$, $E^f_{(110)}$, $E^f_{(111)}$), and vacancy formation energy ($E^f_{vac}$) for Cu and W.

Besides, the equation of state (EOS) curves of the bcc, hcp, dhcp, and fcc phases for Cu and W are shown in Figure 3. $E_0$ represents the ground-state energy of the most stable configuration for Cu (fcc) and W (bcc). The ACE potential accurately predicts the stability order of each phase, i.e., fcc > dhcp > hcp > bcc for Cu and bcc > fcc > hcp for W. ACE predicts $E_{\Delta fcc-bcc}^{Cu(ACE)} = -0.037$ eV/atom and $E_{\Delta fcc-bcc}^{W(ACE)} = 0.470$ eV/atom, which are in excellent agreement with DFT ( $E_{\Delta fcc-bcc}^{Cu(DFT)} = -0.045$ eV/atom, $E_{\Delta fcc-bcc}^{W(DFT)} = 0.471$ eV/atom). In summary, the ACE potential can capture the relative stability of various phases, demonstrating the ability to predict the thermodynamical properties.

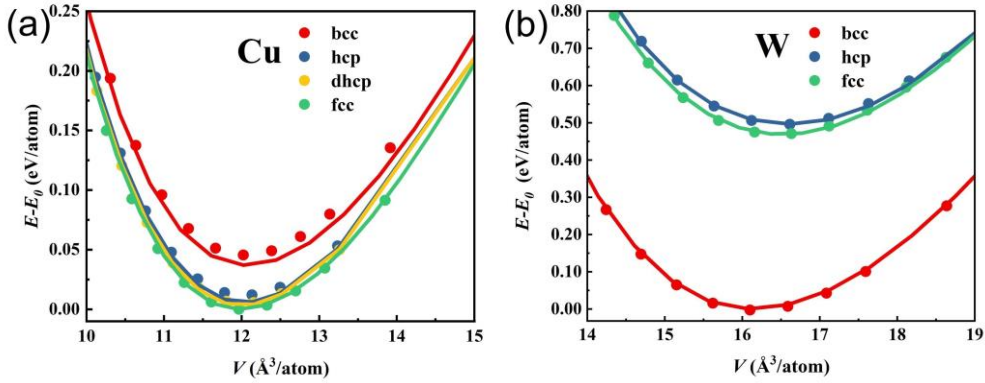

FIG. 3. Low-energy bulk phases of (a) Cu and (b) W. ACE prediction and DFT reference data are plotted in lines and dots respectively.

## B. Phonons

Phonons play a crucial role in describing the mechanical, vibrational, thermal, and dynamic properties of materials [39]. In the phonon spectrum, the highly symmetrical point path has a significant impact on calculating the phonon wave vector, determining the symmetry of the lattice vibration mode, and analyzing the splitting of phonon modes within the crystal structure. The phonon dispersions along high symmetry directions of the Brillouin zone and phonon densities of states for the fcc phase of Cu and the bcc phase of W are shown in Figure 4. The phonon dispersion predicted by ACE is in good agreement with the DFT, especially at low frequencies, which is consistent with the excellent quantitative description of elastic constants.

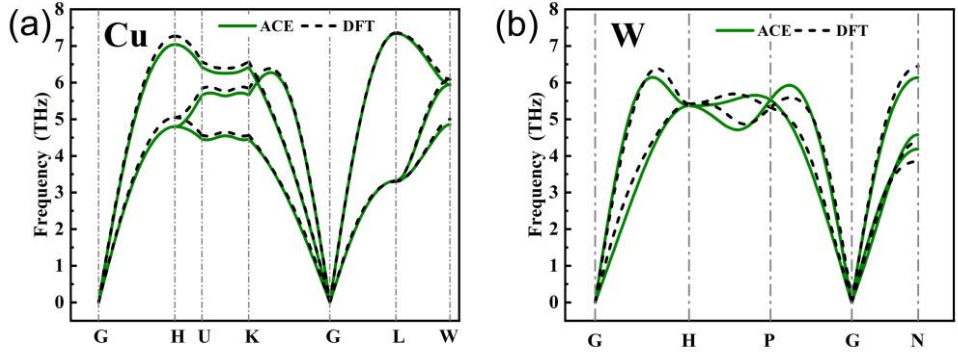

FIG. 4. Phonon dispersion for the ground state (a) fcc structure of Cu and (b) bcc structure of W.

## C. Melting temperature

We simulated the melting behavior of Cu and W using the two-phase simulation method [40]. The melting points calculated via this method are more accurate than those obtained by the single-phase method. In the two-phase simulation method, the initial configuration was constructed using an 8x8x16 supercell, consisting of half of the liquid phase (Cu, 2000K; W, 5000K) and half of the solid phase (Cu, 300K; W, 300K) (see Supplementary Material S2). The initial supercell was equilibrated under the canonical ensemble for 4 ps. The density-temperature curve is obtained according to the density change of the supercell during the simulation process [41], as shown in Figure 5. For both Cu and W, the melting points predicted by the Cu-W ACE potential are in excellent agreement with DFT [19] and experiment [42,43].

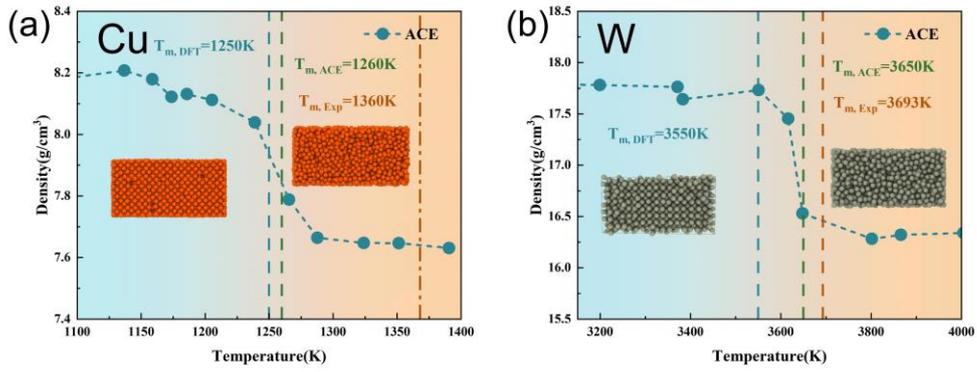

FIG. 5. Relationship between density and temperature for (a) Cu and (b) W.

## D. Coefficients of thermal expansion

Cu-W, as an excellent composite material, is commonly used in electronic packaging materials and heat sinks due to its good thermal conductivity, electrical conductivity, high hardness, and appropriate coefficient of thermal expansion [8]. The lattice parameters as a function of temperature for pure Cu, W, and their Cu-W solid solutions ($L1_2$-$Cu_3W$, $L1_2$-$CuW_3$, and $B_2$-CuW) are shown in Figure 6. With increasing Cu concentration, the lattice parameters gradually decrease for fcc and bcc structures. This agrees well with the positive heat of formation in Cu-W solid solutions, which leads to repulsive interactions between Cu and W atoms within the solid solution [14]. As temperature increases, all lattice parameters exhibit an increasing trend.

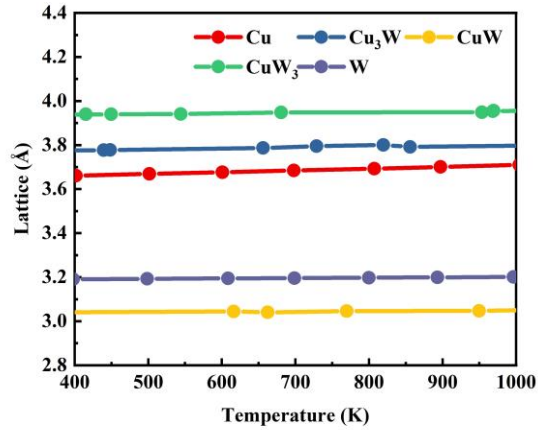

FIG. 6. Relationship between lattice and temperature for Cu, W, and Cu-W solid solutions (Cu$_3$W, CuW, and CuW$_3$).

To investigate the influence of Cu concentration on the coefficient of thermal expansion, we calculated the coefficients of thermal expansion using the lattice constants measured for both fcc (32-atoms) and bcc (16-atoms) structures by varying the Cu concentration. The coefficient of thermal expansion was determined using the lattice constants measured at various temperatures ranging from 300K to 1200K. The NPT ensemble was employed during the simulation under constant zero pressure. The simulation timestep was set as 1.0 fs and the simulations were kept for 10 ps at each temperature step. As shown in Figure 7, W has a smaller coefficient of thermal expansion than Cu due to its stronger atomic bonding. Consequently, with increasing Cu concentration, the coefficient of thermal expansion increases, which is consistent with findings reported in the literature [44]. Therefore, the coefficients of thermal expansion can be adjusted by varying the Cu concentration.

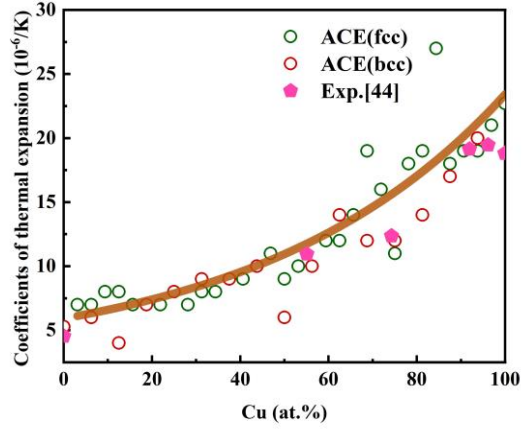

FIG. 7. Coefficients of thermal expansion of CuW alloy with different Cu concentrations.

### E. Diffusion coefficients

Diffusion plays an important role in the mechanical properties, microstructure evolution, and service life of the material, especially at high temperatures. In order to verify the Cu-W ACE potential in describing the diffusion behavior, we calculated the diffusion coefficients of pure Cu and W using Le Claire's nine-frequency model (9-jump model) [45]. The self-diffusion coefficient $D$ can be considered as the product of vacancy concentration $C_v$ and vacancy diffusion coefficient $D_v$ [46]. For pure metals, the vacancy diffusion coefficient $D_v$ is expressed as

$$D_v = f_0 a^2 \omega_0, \tag{4}$$

where $a$ is the lattice parameter, and $f_0$ is the correlation factor [47]. $f_0$ is 0.78146 for fcc structures and 0.72149 for bcc structures. The jump frequency is $\omega_0 = v_0 \exp(-\frac{E_0^m}{k_B T})$, where $v_0$ is the Debye frequency, and $E_0^m$ is the migration barrier. We employed the nudged elastic band method to calculate the energy barriers [48]. The calculation method is given in Supplementary Material S3. We estimated the migration barrier for Cu ($E_0^m = 0.74$ eV) and W ($E_0^m = 1.87$ eV).

The vacancy concentration $C_v$ can be calculated from

$$C_v = \exp[(TS_{v_0}^f - E_{v_0}^f)/k_B T], \tag{5}$$

where $k_B$ is the Boltzmann constant. $S_{v_0}^f$ is vacancy formation entropy, and $E_{v_0}^f$ is vacancy formation energy. The vacancy formation entropy $S_{v_0}^f$ was calculated using the Phonopy code [49]. The calculation was based on the phonon entropies $S_{id}$ and $S_{v_0}$ obtained at $T_0 = 300K$ for two simulated systems. $S_{id}$ is the entropy associated with the perfect crystal, and $S_{v_0}^f$ is entropy for the crystal with one vacancy. The vacancy formation entropy $S_{v_0}^f$ is expressed as

$$S_{v_0}^f = S_{v_0} - (N-1)S_{id}, \tag{6}$$

where $N$ is the number of atoms in the simulation cell with one vacancy (107 and 215 atoms in a simulation cell for Cu and W, respectively). We estimated $S_{v_0}^f = 2.45 k_B$ for Cu and $S_{v_0}^f = 3.78 k_B$ for W. Combined with the $E_{v_0}^f$ calculated in Sec. III.A, the diffusion coefficient is computed and plotted in Figure 8. The slopes of the line predicted by the ACE potential are 14% and 21% lower for Cu and W compared to other calculations [46,50]. However, the trends of the diffusivity of both Cu and W are correctly predicted. The discrepancies between ACE and other calculations can be due to the differences in computational methods and model dimensions.

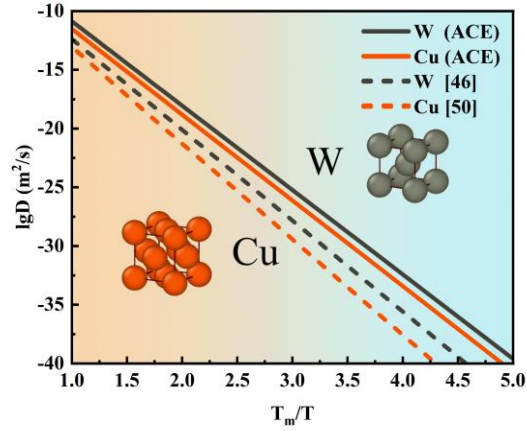

FIG.8. Temperature dependence of the atomic diffusion coefficients in pure metals.

### F. Stacking fault energy

The plastic behavior in metals is mainly governed by dislocations, and the generalized stacking fault energy (GSFE) is closely related to the glide of dislocations. We calculated the GSFE for Cu and W by shifting the basis vector of the corresponding periodic boundaries [51]. As shown in Figure 9, the GSFE curves of {111}<112>, {111}<110> slip systems for Cu and {112}<111>, {112}<111> slip systems for W are calculated by ACE, EAM [12], and DFT [52]. Figures 9(a) and (b) illustrate that ACE is more accurate in describing the unstable stacking fault configurations of Cu than EAM [12]. Figures 9(c) and (d) demonstrate that ACE effectively identifies the precise locations of unstable stacking fault configurations in W. This finding serves to corroborate the accuracy of ACE over EAM [12]. Therefore, the trained ACE potential is reliable for large-scale simulations of plastic deformation.

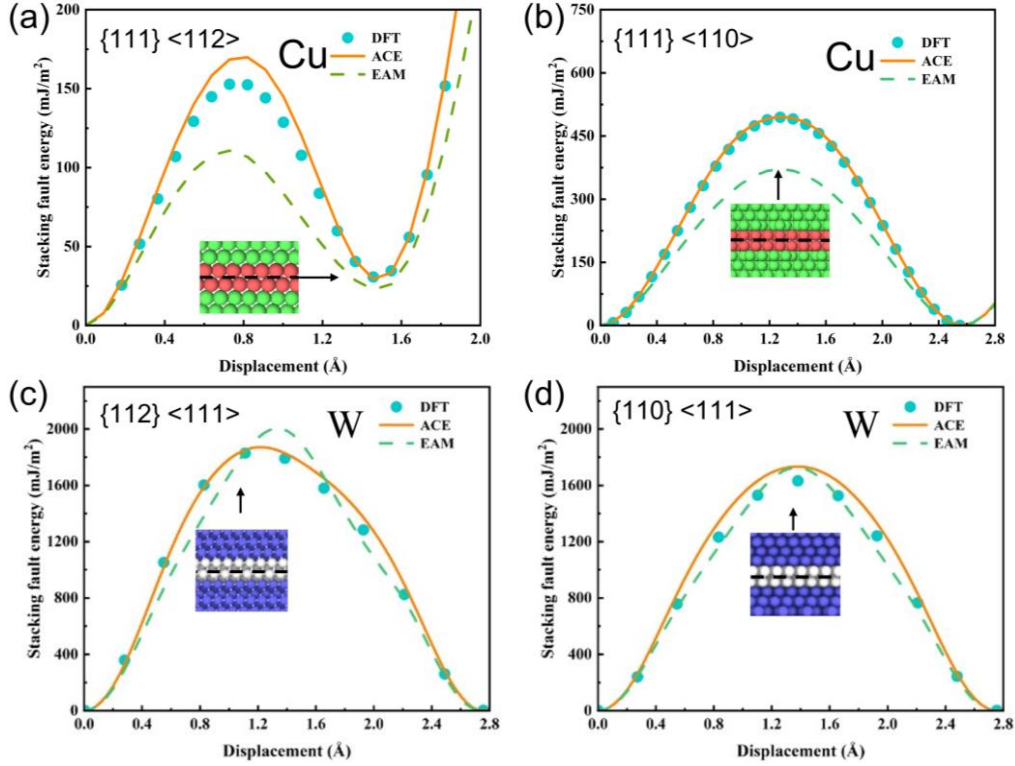

FIG. 9. The generalized stacking fault energy (GSFE) curves of Cu (a) {111}<112>, (b) {111}<110> and W (c) {112}<111>, (d) {110}<111> predicted by DFT [52], ACE, and EAM [12].

### G. Cu-W formation energy

As a typical immiscible system, the solid solution should have a positive formation energy. We calculated the formation energies of the Cu-W solid solution to reveal the phase-separation process. We first validated the performance of ACE on the Cu-W solid system by computing the energy volume curves of three Cu-W solid solutions ($L1_2$-$Cu_3W$, $L1_2$-$CuW_3$, and $B_2$-$CuW$), pure Cu and W. The results are compared against DFT predictions, as shown in Figure 10. The Cu-W ACE potential faithfully reproduces the equilibrium volume and energy of the Cu-W crystalline phases.

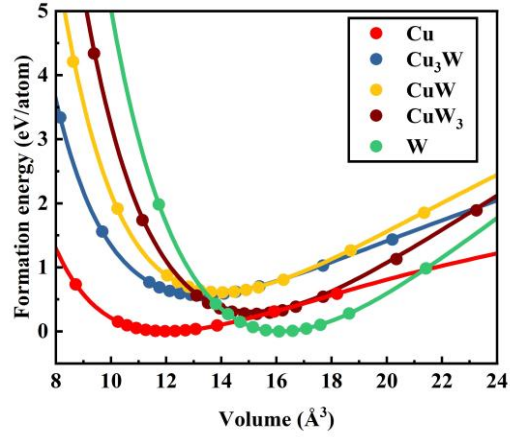

FIG. 10. EOS curves for Cu-W compounds calculated using ACE (lines) compared to DFT (dots).

Based on the energies of the equilibrium state, the convex hull of Cu-W compounds is constructed, as shown in Figure 11. The heat of formation ($\Delta H_f$) for the Cu-W system is calculated according to

$$\Delta H_f^{(Cu_xW_{1-x})} = (E^{(Cu_xW_{1-x})} - xE^{(Cu)} - (1-x)E^{(W)})N_A, \qquad (7)$$

where $N_A$ is the Avogadro constant, and $E^{(Cu_xW_{1-x})}$ is the total energy per formula unit of the Cu-W system. $E^{(Cu)}$ and $E^{(W)}$ are the energies per atom of Cu and W, respectively. Figure 11 shows the heat of formation predicted by ACE and DFT [53]. For the part with Cu concentration less than 50%, the empirical potential function fails to describe the heat of formation for Cu-W solid solutions [14], indicating that the EAM is inaccurate for studying Cu-W systems. On the contrary, the ACE potential is in good agreement with DFT [53] across the full composition range for both bcc and fcc Cu-W solid solutions as well as for some special structures. Therefore, the ACE potential can be effectively utilized for investigating Cu-W systems with low Cu concentration.

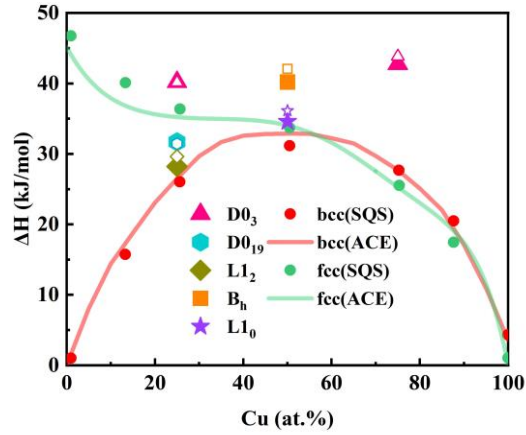

FIG. 11. The convex hull of Cu-W compounds calculated using ACE (open symbols and solid lines) and DFT [53] (solid symbols and dots).

### H. Cu-W interface energy

Experimental observations suggest that fcc/bcc multilayer systems often form interface structures with Nishiyama-Wasserman (NW) and Kurdjumov-Sachs (KS) orientation relations [54-58]. Understanding the thermodynamics of those interfaces is crucial for elucidating the phase separation phenomenon observed in immiscible systems. We calculated the work of separation for the KS and NW interfaces based on the ACE potential and compared the results against DFT predictions (see the Supplementary Material S4). The work of separation is calculated according to

$$W_{sep} = \frac{E^{(W)} + E^{(Cu)} - E^{(tot)}}{A} , \qquad (8)$$

where $E^{(tot)}$ is the total energy of the system, and $A$ is the surface area of the Cu-W interface. $E^{(Cu)}$ and $E^{(W)}$ are the energies of Cu and W surface layers after the removal of the Cu and W layers, respectively. As shown in Figure 12, the Cu-W ACE potential reproduces the work of separation of the Cu-W interface. It is worth noting that the training dataset does not include those interface structures, which further validates the excellent extrapolation performance of the ACE potential.

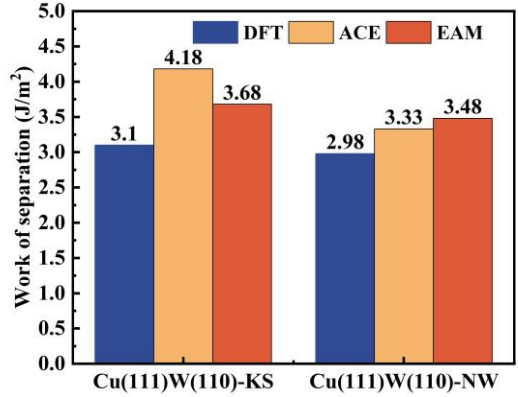

FIG. 12. Work of separation ($W_{sep}$) of Cu-W interfaces predicted by ACE, DFT, and EAM [12] potential.

## I. Phase separation

Nano-composites of phase separating metals are under intensive investigation due to their superior properties, such as high strength [59-62] and radiation resistance [63-65]. The phase separation phenomenon in Cu-W composite materials is prominently observed during the preparation of Cu-W nano-multilayers. However, at the atomic scale, the effects of elemental distribution and concentration on the phase separation process remain unclear.

Based on the Cu-W ACE potential, we performed Monte Carlo Molecular Dynamics (MC-MD) simulations using the Large-scale Atomic/Molecular Massively Parallel Simulator (LAMMPS) package [66,67] to investigate the phase separation of Cu-W nano-multilayers. As shown in Figure 13, W is distributed sinusoidally along the $z$ direction in the bcc solid solution model [68]. The NPT ensemble was employed during the simulation under constant $T = 300\mathrm{K}$ and zero pressure. The simulation timestep was set as 1.0 fs. A hybrid MC-MD scheme was used for the simulation of phase separation using the MC [67] package implemented in LAMMPS [66], which allowed for atomic swapping between different atomic species. The swapping process randomly selected Cu and W atoms within all regions of the bulk and swapped their atomic positions while the total number of each atomic species was fixed. Phase separation simulations were performed for 1 ns. After separation, potential energy convergence was carefully checked to ensure that thermodynamical equilibrium had been reached. The open visualization tool OVITO was used for the visualization of model [69].

We observed a strong tendency of phase separation during the simulations. In the Cu-W immiscible system, the heat of formation for solid solutions is positive, indicating that the Cu and W atoms tend to aggregate into clusters according to thermodynamics. In fact, Figure 13 (b) shows that the layer with a relatively high initial Cu(W) concentration eventually has a higher Cu(W) concentration after MC-MD simulation, indicating that both Cu and W diffuse from their low concentration layer into the high concentration layer. This is consistent with the uphill diffusion phenomenon observed in the experiment [8]. In the simulation, the interlayer is mainly a coherent bcc/bcc interface, which is also in line with the experimental observation [68]. The results with fcc solid solution model are shown in Supplementary Material S5. Interestingly, the initial fcc configuration transformed into a bcc structure, becoming polycrystalline because of the huge energy gap between fcc and bcc phases of W ($E_{\Delta fcc-bcc}^{W(ACE)} = 0.470$ eV/atom).

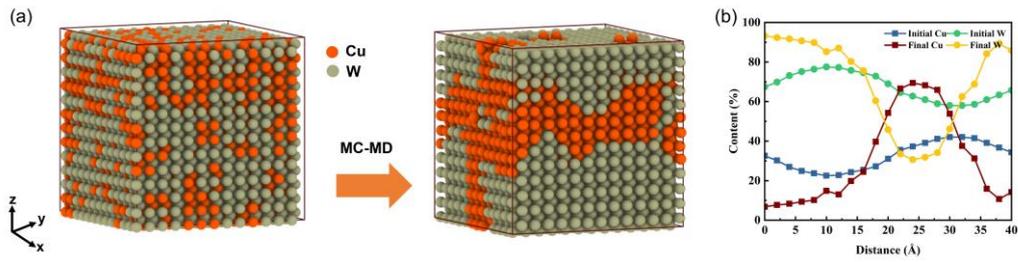

FIG. 13. (a) The initial and final structures during the hybrid MC-MD simulation of the Cu-W system. (b) The distribution of elements along the *z* direction for configurations corresponding to (a).

Furthermore, we created a model with the bcc-initial configuration of 8nm×8nm×12nm to investigate the size effects. As shown in Figure 14 (a), stratification still occurs with a larger model. It can be seen from Figure 14 (b) that the layering period of the Cu-W nano-multilayers is closely related to the distribution of Cu-W elements. This has significant implications for the construction of Cu-W nano-multilayers with special structures. We also investigated the effect of the initial random distribution of elements (see Supplementary Material S6). The solid solution only exhibited element clustering without distinct layer formation, further confirming that the distribution of elements played a crucial role in the formation of the nano-multilayers.

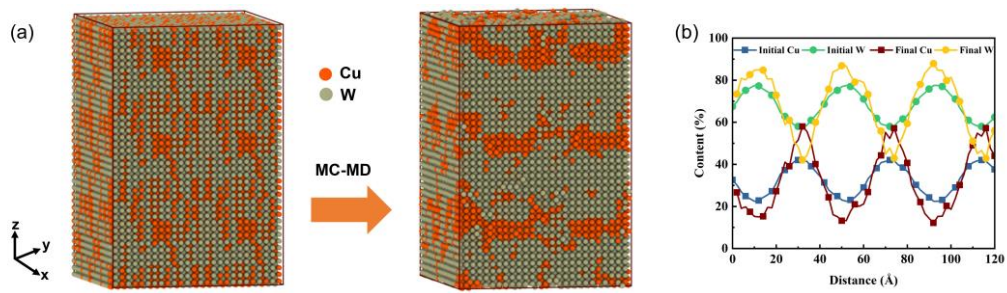

FIG. 14. (a) The initial and final structures during the hybrid MC-MD simulation of the Cu-W system. (b) The distribution of elements along the *z* direction for configurations corresponding to (a).

## IV. Conclusion

In this study, we developed an accurate ML potential for the Cu-W binary system based on the state-of-the-art ACE framework. The developed ACE potential accurately describes the fundamental properties of the Cu-W system with *ab initio* accuracy and is therefore suitable for large-scale atomistic simulations. We comprehensively assessed the accuracy of the ACE potential for thermodynamical properties including the melting point, coefficient of thermal expansion, diffusion coefficient, EOS curves of the Cu-W solid solutions and work of separation for KS and NW interfaces. The phase separation phenomenon of the Cu-W immiscible system was reproduced by performing hybrid MC-MD simulations with the ACE potential. We provided a general Cu-W potential for the investigation of the defects and thermodynamical behaviors of the Cu-W system. The current work also provided a comprehensive DFT database for future benchmarking, development and fine-tuning of interatomic potentials in binary Cu-W and higher order Cu-W-based intermetallic systems. The method of constructing the Cu-W ACE potential can be generalized to other immiscible metal systems as well.

**Authorship contribution statement**

J.H. Pan: Conceptualization, Validation, Investigation, Writing - original draft, Writing - review & editing, Visualization. H.Q. Cheng: Writing -review & editing. G.S. Yan: Writing -review & editing. L. Zhang: Writing -review & editing, Supervision. W.S. Yu: Writing -review & editing, Supervision, Funding acquisition. S.P. Shen: Conceptualization, Supervision, Writing - review & editing, Funding acquisition.

**Acknowledgment**

The authors acknowledge the support of NSFC (Grant No: 12090030).

**Data availability**

The Cu-W ACE potential, DFT database, as well as the simulation workflow used for testing, are made publicly available on GitHub https://github.com/Addition-P/Cu-W-ACE.

**Declaration of Competing Interest**

The authors declare that they have no known competing financial interests or personal relationships that could have appeared to influence the work reported in this paper.

**Supplementary Materials**

Supplementary data related to this article can be found in the Supplementary Materials file.

# Supporting information for "Atomic cluster expansion interatomic potential for defects and thermodynamics of Cu-W system"

Jiahao Pan[1], Huiqun Cheng[1], Gaosheng Yan[1], Lei Zhang[*2], Wenshan Yu[*1], Shengping Shen[*1]

[1]State Key Laboratory for Strength and Vibration of Mechanical Structures, Shaanxi Engineering Laboratory for Vibration Control of Aerospace Structures, School of Aerospace Engineering, Xi'an Jiaotong University, Xi'an 710049, People's Republic of China.

[2]Engineering and Technology Institute Groningen, Faculty of Science and Engineering, University of Groningen, Nijenborgh 4, 9747 AG Groningen, The Netherlands.

[*]corresponding to: lei.zhang@rug.nl (L. Zhang) wenshan@xjtu.edu.cn (W.S. Yu) sshen@xjtu.edu.cn (S.P. Shen)

**S1 The basic properties of Cu and W**

Table SI lists the detailed information regarding the basic properties of Cu and W.

TABLE SI. Basic properties of Cu and W.

| **Copper** | | | |
|---|---|---|---|
| | DFT | ACE | Exp. |
| Lattice(Å) | 3.635 | 3.632 | 3.61[3] |
| Elastic module (GPa) | | | |
| $C_{11}$ | 178 | 182 | 177[4] |
| $C_{12}$ | 123 | 125 | 125[4] |
| $C_{44}$ | 75 | 71 | 81[4] |
| Bulk module (GPa) | 141 | 144 | 142 |
| Vacancy formation(eV) | 1.07[1] | 1.02 | 1.27[5] |

| Surface energy $\gamma_{surf}$ (J/m²) | | | |
|---|---|---|---|
| (100) | 1.51[1] | 1.43 | |
| (110) | 1.57[1] | 1.51 | |
| (111) | 1.36[1] | 1.28 | |
| **Tungsten** | | | |
| | DFT | ACE | Exp. |
| Lattice(Å) | 3.185 | 3.181 | 3.16[6] |
| Elastic module (GPa) | | | |
| $C_{11}$ | 528 | 536 | 532[6] |
| $C_{12}$ | 201 | 202 | 204[6] |
| $C_{44}$ | 152 | 172 | 163[6] |
| Bulk module (GPa) | 310 | 313 | 323[6] |
| Vacancy formation(eV) | 3.27[2] | 3.33 | |
| Surface energy $\gamma_{surf}$ (J/m²) | | | |
| (100) | 4.02[2] | 3.99 | |
| (110) | 3.27[2] | 3.27 | |
| (111) | 3.56[2] | 3.56 | |

## S2 The model for calculating the melting temperature using TPS

Figure S1 shows the models for melting calculation. The model contains 4096 atoms for Cu and 2048 atoms for W. One half of the atoms were kept in the solid state while the other half was heated to a high temperature for 4 ps to ensure complete melting of the unfixed atoms. The temperature was gradually raised to the predicted melting point ($T_m$), establishing an initial configuration of solid-liquid coexistence for each half.

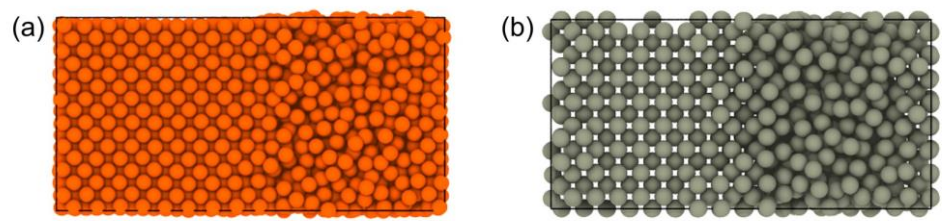

FIG. S1. Configuration of the supercell at the initial state of TPS with 8 × 8 × 16 unit cells for (a) Cu and (b) W. This plot was generated by OVITO [7].

## S3 Migration barrier calculation

The single vacancy migration barriers for Cu and W are shown in Figure S2. The model contains a single vacancy in a Cu (107 atoms) and W (53 atoms) supercell, respectively. The energy barrier was calculated with 11 images in which all atoms of the supercell were allowed to relax using the nudged elastic band method.

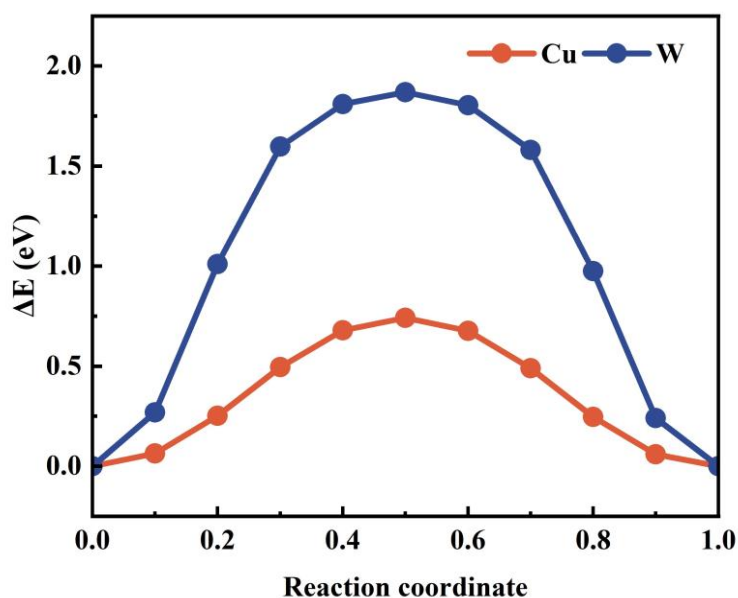

FIG. S2. The calculated migration barriers for Cu and W atoms moving by exchange for the vacancy.

**S4 KS and NW modeling and work of separation calculation**

As shown in Figure S3, we constructed KS and NW containing 114 and 84 atoms using Atomsk [8], respectively. Due to the considerable lattice mismatch between Cu and W, the KS and NW interfaces are incoherent. Table SII lists the detailed information for modeling. The cutoffs for plane wave energies were 500 eV. We used Vaspkit software to generate Γ-centered k-points automatically with a k-mesh spacing of 0.1 Å$^{-1}$

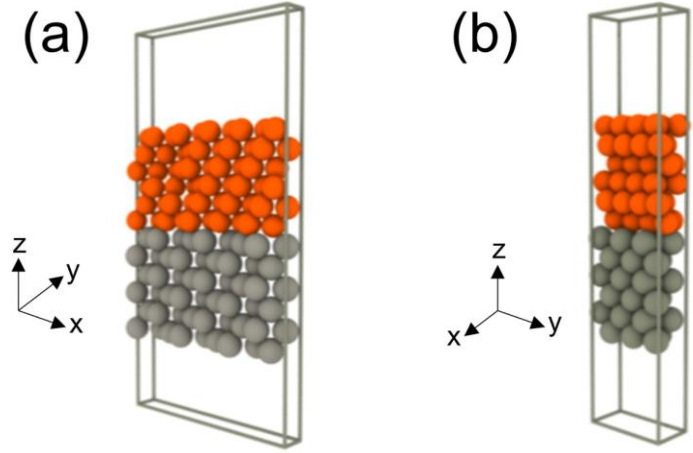

FIG. S3. Configurations of the (a) KS and (b) NW interfaces.

Table SII. Parameters used for constructing KS and NW bicrystals

| Interface | Element | Lattice | Direction | | | Nx | Ny | Nz |
|---|---|---|---|---|---|---|---|---|
| | | | X | Y | Z | | | |
| KS | Cu | 3.727 | [11-2] | [1-10] | [111] | 5 | 1 | 2 |
| | W | 3.106 | [1-12] | [-111] | [110] | 3 | 1 | 3 |
| NW | Cu | 3.503 | [11-2] | [1-10] | [111] | 1 | 4 | 2 |
| | W | 3.303 | [-110] | [00-1] | [110] | 1 | 3 | 3 |

**S5 Phase separation with an initial fcc solid solution model**

Figure S4 plots the phase separation predicted by the ACE potential with an initial fcc model. The simulation result reveals a pronounced phase separation after the simulation. Interestingly, the initial fcc configuration transformed into a bcc structure, becoming polycrystalline due to the high energy gap ($E_{\Delta fcc-bcc}^{W(ACE)} = 0.470$ eV/atom) between bcc and fcc phases. The fcc W is unstable with high potential energy, which drives the configuration to transform into bcc.

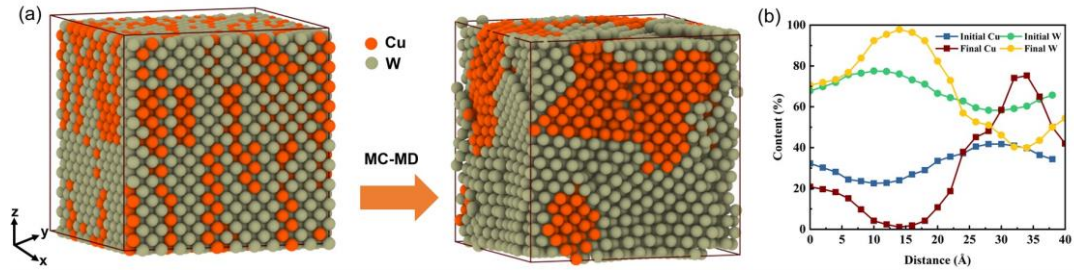

FIG. S4. (a) The initial and final structures during the hybrid MC-MD simulation of the Cu-W system. (b) The distribution of elements along the z direction for configurations corresponding to (a).

## S6 Randomly distributed elements

Figure S5 plots the phase separation predicted by ACE with randomly distributed elements. To explore the effect of the distribution of elements, we built a model with randomly distributed elements, and the rest of the settings were the same as in the main manuscript. However, it only exhibited element clustering without distinct layer formation, further confirming that the distribution of elements played a crucial role in the formation of the nano-multilayers.

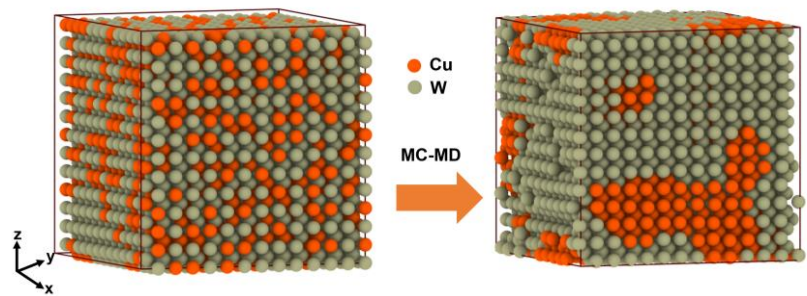

FIG. S5. The initial and final structures during the hybrid MC-MD simulation of the Cu-W system.